\documentclass[useAMS]{mn2e}
\usepackage{graphicx,times,aas_macros}

\title[XCS: X-ray source identification using the SDSS]{XMM
Cluster Survey: X-ray source identification
using the Sloan Digital Sky Survey} \author[K. R. Land et al. (The
XCS collaboration)]{Kate R. Land,$^{1,2}$ Robert C. Nichol,$^3$
Michael Davidson,$^4$ Kivanc Sabirli,$^3$ \cr A. Kathy Romer,$^{1,3}$
Andrew R. Liddle,$^{1}$ Chris A. Collins,$^5$ Scott T. Kay,$^{1}$ \cr
Robert G. Mann,$^{4,6}$ Pedro T. P. Viana,$^{7,8}$ and Michael
J. West$^9$ \cr (The XCS Collaboration)\\ 
\vspace*{-6pt} {\small \em $^{1}$Astronomy Centre, University of Sussex, 
Brighton BN1 9QH, U.~K.}\\
\vspace*{-6pt} {\small \em $^{2}$Theoretical Physics, The Blackett Laboratory, 
Imperial College, Prince Consort Road, London SW7 2BZ, U.~K.}\\
\vspace*{-6pt} {\small \em $^{3}$Department of Physics, Carnegie Mellon 
University, 5000 Forbes Ave., Pittsburgh, PA-15217, U.~S.~A.}\\
\vspace*{-6pt} {\small \em $^{4}$Institute for Astronomy, University of 
Edinburgh, Blackford Hill, Edinburgh, EH9 3HJ. U.~K.}\\
\vspace*{-6pt} {\small \em $^{5}$Astrophysics Research Institute, Liverpool John 
Moores University, Twelve Quays House, Egerton Wharf, Birkinhead CH41 1LD, 
U.~K.}\\
\vspace*{-6pt} {\small \em $^{6}$National E-Science Centre, South College 
Street, Edinburgh EH8 9AA, U.~K.}\\
\vspace*{-6pt} {\small \em $^{7}$Centro de Astrof\'{\i}sica da Universidade do 
Porto, Rua das Estrelas, 4150-762 Porto, Portugal}\\
\vspace*{-6pt} {\small \em $^{8}$Departamento de Matem\'atica Aplicada da 
Faculdade de Ci\^encias da Universidade do Porto, Rua do Campo Alegre 687, 
4169-007 Porto, Portugal}\\
\vspace*{-6pt} {\small \em $^{9}$Department of Physics and Astronomy,
University of Hawaii, 200 W. Kawili Street, Hilo, Hawaii 96720,
U.~S.~A.  }}

\begin{document}
\journal{Preprint astro-ph/0405225} 

\date{\today}

\maketitle

\label{firstpage}

\begin{abstract}
The XMM Cluster Survey (XCS) is predicted to detect thousands of
clusters observed serendipitously in XMM-Newton pointings. We investigate 
automating optical follow-up of XCS cluster
candidates using the Sloan Digital Sky Survey (SDSS) public
archive,
concentrating on 42 XMM observations that overlap the First Data Release
of the SDSS.  Confining ourselves to the inner 11 arcminutes of the
XMM field-of-view gives 637 unique X-ray sources across a
$3.09\,{\rm deg^2}$ region with SDSS coverage. The log$N$--log$S$
relation indicates survey completeness to a flux limit of around
\mbox{$1\times10^{-14} \, {\rm erg} \, {\rm s}^{-1} \, {\rm cm}^{-2}$}
(in the 0.5-2.0 keV band).  

We have used the SDSS data in a variety of
ways. First, we have cross-correlated XMM point sources with SDSS
quasars, finding 103 unique matches from which we determine a 90\% confidence
matching radius of 3.8 arcseconds.  Using this matching radius we make
immediate identifications of roughly half (214 of 520) of 
all XMM point sources as quasars (159) or stars (55). These objects
will be a powerful resource for non-cluster studies. Second, we have
estimated the typical error on SDSS-determined photometric cluster
redshifts to be $\simeq 5\%$ for relaxed systems, and
$\simeq 11\%$ for disturbed systems, by comparing photometric
redshifts to spectroscopic redshifts for eight previously-known
clusters ($0.183<z_{{\rm spec}}<0.782$). The measured level of error
for disturbed systems may be problematic for X-ray and
Sunyaev--Zel'dovich cluster surveys relying on photometric redshifts alone.
Third, we use the False Discovery Rate methodology to
select 41 XMM sources (25 point-like, 16 extended) statistically
associated with SDSS galaxy overdensities. Of the 16 extended sources,
5 are new cluster candidates and 11 are previously-known clusters
($0.044<z<0.782$). There are 83 extended X-ray sources not
associated with SDSS galaxy overdensities, which are our strongest candidates 
for new high-redshift clusters.
We highlight two
previously-known $z>1$ clusters that were rediscovered as extended
sources by our wavelet-based detection software. 

These results show that the SDSS can provide
useful automated follow-up to X-ray cluster surveys, especially with
regard to the rejection of point-source contamination, but cannot provide all 
the optical follow-up necessary
for X-ray cluster surveys. Additional
telescope resources must be employed to ensure their timely completion.
\end{abstract}


\section{Introduction}

The number density of galaxy clusters has the potential to provide
important tests of the cosmological model (e.g.~Evrard 1989; Henry \&
Arnaud 1991; Oukbir \& Blanchard 1992; White, Efstathiou \& Frenk 1993;
Eke, Cole \& Frenk 1996; Viana \& Liddle 1996, 1999; Henry 1997; Eke
et al.~1998). We describe here an ongoing effort to construct a new
catalogue of clusters for cosmological studies, using X-ray data
available in the XMM-Newton archive. This effort, known as the XMM
Cluster Survey (XCS), has been described previously by Romer et
al.~(2001). By the end of the XMM mission, XCS is expected to include
thousands of clusters of galaxies out to redshifts well beyond one. Optical 
follow-up of
cluster candidates is one of the key challenges facing both X-ray
and Sunyaev--Zel'dovich (e.g.~Carlstrom, Holder \& Reese 2002)
surveys for distant clusters.  In the past this was achieved using
dedicated follow-up programmes that took many years to complete
(e.g.~Romer et al.~2000; Perlman et al.~2002; Mullis et al.~2003;
Gioia et al.~2003). However, with the recent emergence of the
``Virtual Observatory'' (Szalay \& Gray 2001) and datasets such as the
Sloan Digital Sky Survey (SDSS; York et al.~2000), it should be
possible to reduce the time spent on optical follow-up work. A first
application of this to X-ray cluster surveys has been made by
Schuecker, B\"ohringer \& Voges (2004) using SDSS and ROSAT All Sky
Survey (RASS) data.  

This paper presents preliminary results from the
matching of SDSS data with the XCS. In Section~\ref{data}, we discuss
the XMM data and outline our methods for
analyzing it and determining the extent of X-ray sources. 
Section~\ref{s:sdss} describes the SDSS data we use for optical identifications; 
we use this to discuss positional accuracy of the source matching, associations 
with overdensities of optical galaxies, and photometric redshifting.
In Section~\ref{comb}, we analyze the properties both of matched sources and of 
cluster candidates detected only in one of the wavebands.  Throughout
this paper, we assume a concordance flat cosmology ($\Omega_{{\rm m}}=0.3$,
$\Omega_{\Lambda}=0.7$, $h=0.75$) with a cosmological constant. Fluxes
are given in the ($0.5$--$2.0$) keV band unless otherwise stated.

\section{X-ray Data Analysis}
\label{data}

\begin{figure}
\begin{center}
\includegraphics[angle=0,width=0.46\textwidth ]{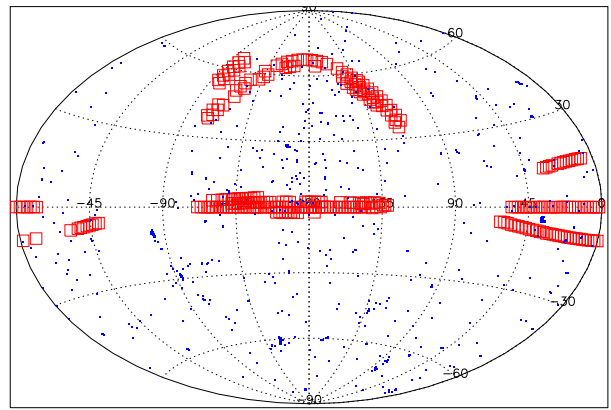}
\caption{The 766 XMM pointings processed in the XCS project as 
of April 2003 are shown as blue points (equatorial projection). The
SDSS-DR1 spectroscopic plates are indicated by red boxes. In each
case, the symbol size indicates actual sky
coverage.\label{radecplots}}
\end{center}
\end{figure}

The XCS will process {\em all} EPIC imaging data available in the XMM
archive. As of April 2003, 766 XMM pointings had been reduced using
the XCS pipeline --- see Figure~\ref{radecplots}. For this paper
however, we restrict ourselves to 42 XMM pointings that lie within the
SDSS first data release (DR1) spectroscopic area (Abazajian et
al.~2003, henceforth A03). The SDSS DR1 comprises 2099 square degrees
of five-band \mbox{(u, g, r, i, z)} imaging data, and 186,240 spectra
of galaxies, quasars, stars selected over 1360 square degrees of the
imaging area.  In Table~\ref{pointings}, we
present the observation identification number (column 1) and
coordinate (column 2) of each pointing, together with the designated
target name and classification (column 3). Where available, we list in
column 4 the target redshift according to the
NED database.\footnote{http://nedwww.ipac.caltech.edu/} We list the
number of sources detected (Section~\ref{sode}) within an off-axis
distance $\theta<11$ arcminutes in column 5. For pointings with only
partial SDSS coverage (indicated in column 8), we also give the number
of $\theta<$11 arcminutes sources with SDSS overlap in parentheses;
there are 50 sources detected in one or more of the pointings that do
not overlap with the SDSS DR1. The maximum exposure time in seconds,
after flare rejection (Section~\ref{xmm}), is given in column 6. The
number of square XMM image pixels ($5.1$ arcseconds on a side) used to
calculate the area of the survey for the log$N$--log$S$ relation is
given in column 7. Pointings with $\ge 52,000$ active pixels in
Table~\ref{pointings} (18 of the 42 pointings) correspond to Full
Frame Mode pointings with almost complete XMM/SDSS DR1 overlap and
where no masking of the XMM field-of-view was necessary.

\begin{table*}
\caption{The 42 XMM Pointings with SDSS DR1 overlap. Certain
pixels were not included when the survey area was calculated for the
log$N$--log$S$ relation: 1) pixels surrounding an off-axis extended
source, 2) pixels surrounding an extended target, 3) pixels not read
out during small-window mode observations.}  
\label{pointings}
\begin{tabular}{lcl|lcrr|l}\hline\hline
Obs. ID & J2000 & Target (Type)  & $z$ & Sources & Time & Pixels & Notes\\ 
\hline
     0002940401&  03 41 16.2, -01 18 58.6& NGC1410 (AGN)          & 0.025 & 7 
(2)  & 7094   & 20942 & Partial SDSS overlap\\
     0012440101&00 56 20.1, -01 19 30.6&  LBQS0053-0134 (QSO)     & 2.063  & 16 
(2) & 28327  &  5175 &  Partial SDSS overlap, 1\\
     0042341301& 23 37 46.7,  00 15 35.1& RXCJ2337.6+0016 (Clus)  & 0.273  & 14 
& 13132  & 45836 & 1, 2 \\
     0044740201& 11 50 43.2,  01 44 17.7& Beta Vir (F Star)       &        & 32 
(25) & 47724  & 52218 & Square hole in SDSS around target\\
     0051760101& 12 46 36.3,  02 20 25.1& PG 1244+026 (AGN)       & 0.048  & 9  
& 12035  & 52222 &  \\
     0056020301& 02 56 37.9,  00 04 59.7&RX J0256.5+0006 (Clus)   & 0.360  & 18 
& 20664  & 52153 &  \\
     0060370101& 15 43 56.5,  54 00 48.2& SBS 1542+541 (QSO)      & 2.371  & 12 
& 8331   & 52171 &  \\
     0060370901& 15 43 55.7,  54 00 43.4& SBS 1542+541 (QSO)      & 2.371  & 25 
& 25098 &       52160 &  \\
     0065140101& 00 41 48.1, -09 23 10.7& Abell 85 (Clus)         & 0.044  & 9  
& 12380 &       37076 &  2\\
     0065140201& 00 42 36.8, -09 42 10.1& Abell 85 (Clus)         & 0.044  & 12 
& 12379 &       52185 &  \\
     0066950301& 23 18 10.2,  00 17 04.9&  NGC 7589 (Galaxy)       & 0.0298 & 5  
& 10833 &       52197 &  Missing stripe in SDSS\\
     0066950401& 23 18 21.5,  00 14 59.6&NGC 7589 (Galaxy)       & 0.0298 & 9  & 
12360 &       52198 &  \\
     0081340801& 12 13 44.8,  02 50 22.6& IRAS12112+03 (Galaxy)   & 0.072  & 22 
& 22295 &      52210 &  \\
     0084230401&  01 52 47.3,  00 59 42.1& Abell 267 (Clus)        & 0.230  & 22 
& 17122  &   47177 &  2\\
     0084230601& 9 17 54.9,  51 41 57.1& Abell 773 (Clus)        & 0.217  & 11  
& 14140 &     42722 &  2\\
     0085640201& 9 35 47.1,  61 22 44.0& UGC 05101 (AGN)         & 0.0394 & 36 & 
33340 &     52154 &  \\
     0090070201& 00 43 25.2,  00 50 16.1& UM 269 (QSO)            & 0.3084 & 24 
& 20204 &     40104 &  1\\
     0092800101& 08 31 37.8,  52 46 52.8& APM 08279+5255 (QSO)    & 3.870  & 19 
& 16520 &     52159 &  \\
     0092800201& 08 31 46.5,  52 43 41.4& APM 08279+5255 (QSO)    & 3.870  & 47 
& 72289 &    52132 &  \\
     0093030101& 13 11 28.3, -01 18 49.6& Abell 1689 (Clus)       & 0.1832 & 34 
& 36996 &     52213 &  \\
     0093060101& 12 04 26.4,  01 55 59.1& MKW4 (Clus)             & 0.0200 & 13 
& 14090 &     47187 &  2\\
     0093200101& 12 58 46.9, -01 43 15.3&   Abell 1650 (Clus)       & 0.0845 & 
24 & 37576 &     39138 &  2\\
     0093630101& 02 41 00.0, -08 14 10.4& NGC 1052 (AGN)          & 0.0049 & 20 
& 15446 &     52154 &  \\
     0093641001& 01 43 07.6,  13 37 32.3& NGC 660 (AGN)           & 0.0028 & 8  
& 11040 &     52600 &  \\
     0094800201& 11 40 33.8,  66 09 55.8& MS1137.5+6625 (Clus)    & 0.7820 & 27 
& 35734 &     52140 &  \\
     0101640201& 01 59 44.8,  00 24 46.8& Mrk 1014 (QSO)          & 0.1630 & 18 
& 9732  &    40071 &  3\\
     0102040501& 14 29 08.6,  01 15 30.2& Mrk 1383 (AGN)          & 0.0865 & 3  
& 3228  &    40094 &  3\\
     0108460301& 23 54 03.4, -10 23 1.20& Abell 2670 (Clus)       & 0.0762 & 16 
& 17298 &     41311 &  1\\
     0108670101& 10 23 38.2,  04 13 00.0& ZwCl 1021.0+0426 (Clus) & 0.291  & 43 
& 53020 &     46257 &  1\\
     0110990201& 12 27 20.0,  01 27 42.1& HI1225+01 (Galaxy)      & 0.0043 & 11 
& 28566 &     52218 &  \\
     0111190701& 12 42 46.5,  02 42 53.8& NGC 4636 (Galaxy)       & 0.0031 & 35 
& 58547 &     37225 &  1\\
     0111200101& 02 42 35.5,  00 00 04.3& NGC 1068 (Galaxy)       & 0.0038 & 29 
(28) & 36128 &     48336 &  Small hole in SDSS data, 1 \\
     0111200201& 02 42 35.6,  00 00 05.4& NGC 1068 (Galaxy)       & 0.0038 & 33 
(32) & 32662 &     48334 &  Small hole in SDSS data, 1\\
     0113040801&  01 20 33.1, -10 56 06.5& C2001                   &        & 9 
(4)  & 8407  &    14447 & Partial SDSS coverage\\
     0113040901& 01 18 41.1, -10 39 49.8& C2001                   &        & 5  
& 9794  &    52240 &  Proton flaring particularly bad\\
     0126700201& 12 29 04.3,  02 00 06.9& 3C 273off-1.5min (QSO)  & 0.1583 & 8  
& 25098 &     47223 &  2\\
     0126700501&  12 29 10.1,  02 02 44.2& 3C 273off+1.5min (QSO)  & 0.1583 & 2  
& 9511  &    48400 &  2\\
     0129350201&  03 36 42.2,  00 36 31.5& HR1099 (G star)         &        & 3 
(1)  & 6374  &    47109 &  Large section missing in SDSS, 2 \\
     0134540101& 03 36 51.9,  00 34 16.4& HR1099 (G star)         &        & 13 
(10) & 22869 &      47108 &  Large section missing in SDSS, 2\\
     0134540401& 03 36 41.9,  00 36 28.0& HR1099 (G star)         &        & 11 
(6) & 11546 &      35687 & Large section missing in SDSS, 3\\
     0134540601&  03 36 42.2,  00 36 21.5& HR1099 (G star)         &        & 24 
(17) & 35100 &      45792 & Section missing in SDSS data, 2\\ 
     0137551001& 12 29 05.8,  02 04 23.4& 3C 273 (QSO)            & 0.1583 & 2  
& 8963  &     35744 &  3\\\hline\hline
\end{tabular}
\end{table*}

\subsection{XMM data reduction and flare cleaning}
\label{xmm}

The raw XMM photon data was obtained from the data archive at
XMM-Newton Science Operations
Centre\footnote{http://xmm.vilspa.esa.es/\label{xmmpage}} and reduced
with the February 2003 version of the XCS pipeline. This pipeline uses
the XMM Science Analysis System (SASv5.4.1) and a single set of
Current Calibration Files (CCF) for mission specific data processing.
In the February 2003 version, only the data from the EMOS1 camera were
used (subsequent versions generate merged images from the EMOS1, EMOS2
and EPN cameras).  A data summary file is generated and, via the
Observation Data Files (ODF) access layer, the completeness of each
set is checked. The event linearization for each imaging mode is
processed using SAS tools \texttt{emproc} and \texttt{epproc} for EMOS
and EPN cameras respectively.

\begin{figure*}
\begin{center}
\includegraphics[angle=0,width=1.0\textwidth ]{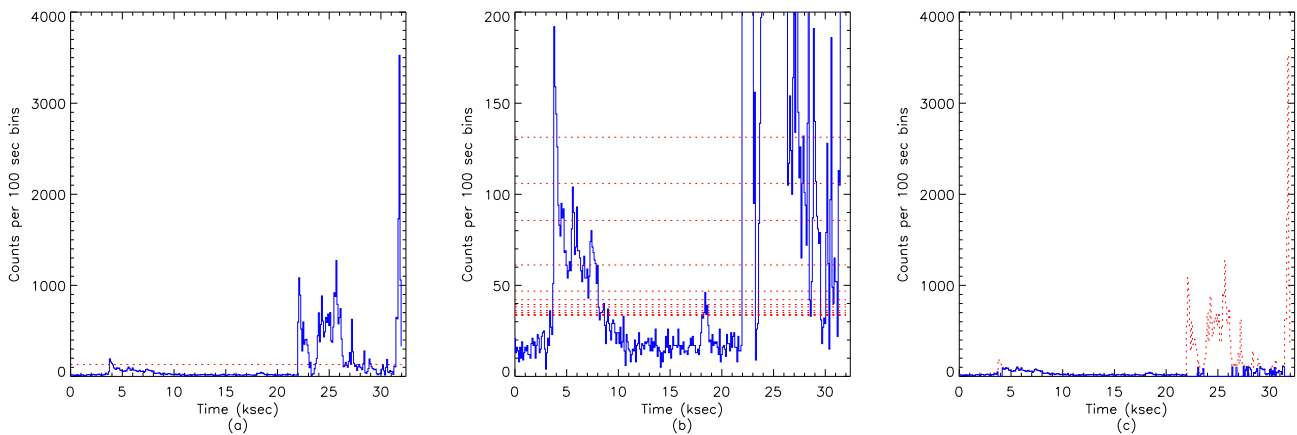}
\caption{This figure shows an example of the high-energy filtering
method we used for each pointing. The data are from observation
0060370901. The lightcurves are generated for events with energies $>$
10 keV with 100 sec binning. (a) The initial light curve is in
blue/solid, and the mean of the uncut light curve is in
red/dashed. (b) The same with reduced vertical scale. The decrease in
the mean of the light curve using successive 3$\sigma$ cut cleaning
can be seen from the mean of the remaining exposure which is in
red/dashed. (c) The original high-energy light curve, now with the
filtered flaring regions shown as red/dashed and the remainder as
blue/solid.\label{flares}}
\end{center}
\end{figure*}

The resulting event lists were checked for temporal filtering to clean
up the high-energy ($>$ 10 keV) flaring periods due to solar protons,
see below.  A detailed study of the XMM background can be found in
Lumb et al.~(2002) and Read \& Ponman (2003).  The flare
cleaning method we used is similar to that suggested by
Marty et al.~(2003) and involves an iterative 3$\sigma$
clipping. An example of this method can be seen in
Figure~\ref{flares}. For each
event list, we filtered the sets with (PATTERN$<=$12).  All sets were
also filtered for energy range 10keV$\le$PI$\le$15keV and the light
curves were generated with 100 sec binning (where PI is ``pulse height
invariant channel''). The mean and standard deviation were
calculated in the counts per bin, and successive 3$\sigma$ cuts of the
high-count periods were applied until the change in the mean was
negligible. Good time intervals were generated using the last mean and
the standard deviation with a last 3$\sigma$ cut and used in filtering
of the previously-prepared raw set with the aforementioned pattern
selection and \mbox{(FLAG == 0)}. Although this method of cleaning
does not give a perfect filtering for all sets, it can be automated to
clean thousands of pointings at once, as required for the
XCS. Considering the varying quiescent mean in the high-energy
background (Pratt \& Arnaud 2002; Lumb et al.~2004),
it is superior to defining a fixed cut for filtering.

After flare cleaning, the products used for source detection were
generated: Detector masks, images and exposure maps in two energy
bands ([0.5--2.0] and [2.0--10.0] keV) with $5.1$ arcsecond square pixels
binning. Exposure maps include spatial quantum efficiency, filter
transmission, mirror vignetting, and field-of-view information. The
image products were then passed into the source detection algorithm. 

\subsection{Source detection}
\label{sode}

We identify sources detected in XCS fields using a wavelet-based
algorithm. The key advantage of wavelets is the natural separation
achieved between small- and large-scale variations in the data,
e.g.~a large-scale slowly-varying background will be filtered out of
the small-scale signal, and therefore will not influence the detection
of such sources [see also Vikhlinin et al.~(1998), Romer et al.~(2000) and
Freeman et al.~(2002)].

For the work presented herein, the source detection proceeded as
follows.  First, we performed a wavelet transform of the 2D XMM image using
the \emph{\`a trous} algorithm (e.g.~Holschneider et al.~1989; Shensa 1992)
to produce, for each spatial scale, a 2D image of wavelet
coefficients from which we must remove noise, i.e.~determine
which coefficients are statistically significant.  One complication
here is the small-number statistics for the photons in the X-ray
background of our images. To accommodate this problem, we have adopted
the ``wavelet histograms'' approach of Starck \& Pierre (1998) as
implemented in their MR/1 software
package,\footnote{http://www.multiresolution.com} in particular the
task \texttt{mr\_detect}. Briefly, the software first identifies all
pixels in a wavelet coefficient image that are above a confidence
level, specified by the user, after modelling the low count rate
Poisson noise in the images. These significant pixels in each wavelet
plane are connected to form the ``segmentation image'' at that
scale. A set of connected wavelet coefficients is known as a
\emph{structure} and structures within different wavelet planes are
connected to form \emph{objects} by way of the \emph{interscale
relation}. A structure $S_{j}^{1}$ at scale $j$ is said to be
connected to a structure $S_{j+1}^{2}$ at scale $j+1$ if $S_{j+1}^{2}$
contains the pixel in $S_{j}^{1}$ with the maximum wavelet
coefficient. In this way objects can be identified in wavelet-transform space 
and can be divided into sub-units of the main source
(i.e.~de-blended). The objects are then reconstructed
iteratively using a 2D Gaussian and the counts associated with each
object computed.  The local exposure time gives the rate in counts per
second. We then use energy conversion factors to obtain the flux,
depending on the absorption factor $n_{{\rm H}}$ and the type of
filter used.  The fluxes of our sources were calculated with energy
conversion factors (ECFs) generated as recommended by Osborne (2003), using the 
canned response matrices released on
2003-01-29 and ancillary response files generated for each filter type
on the optical axis for the pattern selection specified in
Section~\ref{xmm}. To take the effect of the column density of neutral
hydrogen in the direction of each XMM pointing into account, we
generated tables of ECFs for a subset of possible $n_{{\rm H}}$
values. The $n_{{\rm H}}$ values of the pointings were obtained using
one of the FTOOLS,
nH,\footnote{http://heasarc.gsfc.nasa.gov/lheasoft/ftools/heasarc.html}
which uses the HI map by Dickey \& Lockman (1990) as a basis.

\begin{figure}
\centering
\includegraphics[width=8.3cm]{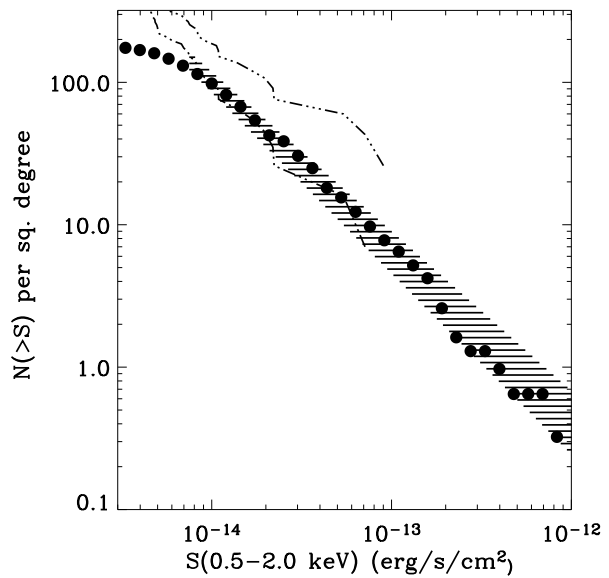}\\
\caption{The log$N$--log$S$ relation (black dots) for XCS sources 
in a $3.09 \,{\rm deg^2}$ region overlapping the SDSS DR1 area.  The
hatched region is the observed range taken from Baldi et al.~(2002).
The dot-dashed lines are error bounds for the XMM log$N$--log$S$ in
the HDF--North region (Miyaji et al.~2003).}
\label{logn}
\end{figure}

When no cut in off-axis distance is applied, we detect 1121 X-ray
sources within the 42 XMM pointings presented in
Table~\ref{pointings}. This number includes duplicate detections (in
overlapping pointings), the primary (usually on-axis) target of the
respective pointing, and regions without SDSS DR1 coverage. We define a
duplicate detection to be one that falls with 5 arcseconds of a source
detected in a different pointing.  Throughout this paper, we
concentrate on those 740 sources that lie within $\theta=11$
arcminutes of the pointing center, i.e.~where the XMM PSF is
well-behaved.

\begin{figure*} 
\begin{center}
\includegraphics[angle=0, width=0.8\textwidth]{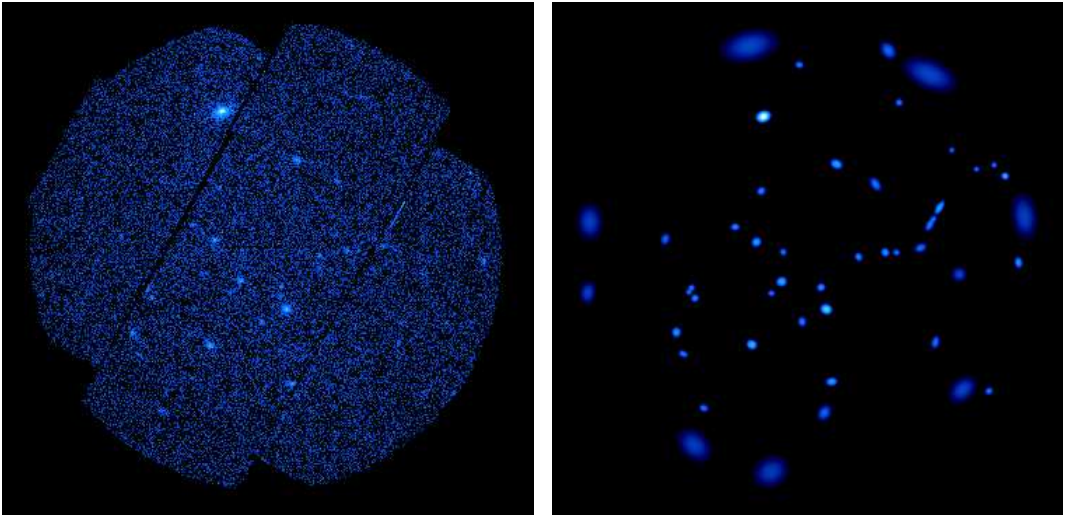}
\caption{On the left we present a sample XMM image, and on the right
we show the location and size of sources detected by the \emph{\`a
trous} algorithm.\label{recon_objects}}
\end{center}
\end{figure*}

In Figure~\ref{logn} we present the $\theta<11$
arcminutes log$N$--log$S$ relation for the 42 pointings in
Table~\ref{pointings}. To create this figure, we have removed
duplicate detections, primary targets and sources in regions without
SDSS DR1 overlap. We determined that there is a unique overlap area of
$3.09\,{\rm deg^2}$ between the 42 pointings and the SDSS DR1. This
area has been corrected for pointing overlaps, partial SDSS cover
and any $\theta<11$ arcminutes regions masked (e.g.~due to the
presence of an extended target) or missing (e.g.~because the
observation was carried out in small-window mode) from the XMM field-of-view.
We did not correct for the different sensitivities of the individual
pointings when constructing the log$N$--log$S$ relation, but note that the
exposure times are similar for most pointings and so this should not
be a significant problem.  Our log$N$--log$S$ relation is in excellent
agreement with the published XMM log$N$--log$S$ relation of Baldi et
al.~(2002) to at least a flux limit of $1\times10^{-14} \, {\rm erg}
\, {\rm s}^{-1} \, {\rm cm}^{-2}$, which indicates that our parent
catalogue is complete to this flux limit. Figure~\ref{logn} demonstrates
the quality of the XMM image processing pipeline.  We do not make a
fit to the log$N$--log$S$ relation, or attempt to model the
incompleteness at fainter flux limits, as we are primarily interested
in finding X-ray clusters of galaxies.

\subsection{Extent classification}
\label{extent}

The next step is to classify the detected sources as extended or point-like. The 
XCS will use X-ray extent as the primary means by which to
select cluster candidates; assuming no size evolution for clusters
with redshift, we expect clusters to be resolved by XMM at any
redshift (Romer et al.~2001).  The challenges here are
two-fold. First, the point-spread function (PSF) of XMM varies across
the satellite field-of-view; this can be seen in
Figure~\ref{recon_objects}. Secondly, over 80\% of all X-ray sources
are point sources (AGNs, quasars, stars etc.), and therefore any
misclassification of X-ray point sources as extended sources will
swamp the underlying cluster population leading to laborious optical
follow-up.  Therefore, any extended source detection algorithm must
address these two issues, as well as being sensitive to true extended
sources. The completeness of any extended source catalogue will need
to be determined via extensive simulations, as in Adami et al.~(2000)
and Burke et al.~(2003).

For the work presented herein, we used a preliminary extended source
detection algorithm based on the ratios of the wavelet coefficients
for each source as calculated at the different wavelet scales. For
example, the wavelet coefficients of point-source objects should be
relatively higher at the smallest wavelet scales, and smaller at the
larger scales; the reverse would be true for an extended source. We
use the square of the modulus of the coefficients to provide a measure
of the wavelet power at that scale and use the ratios of different
scales to classify the sources. For our data, we have four wavelet scales
(plus one low-resolution residual scale which is not used) and
therefore we can calculate six wavelet ratios (not all independent) as
\[
\hspace*{3cm}
R_{ij} = \frac{C_i^2}{C_j^2}, \quad i<j \,, \nonumber
\]
where $C_i$ denotes the wavelet coefficient at the pixel closest to the
source position at one of the four wavelet scales available to us.  

\begin{figure}
\begin{center}
\includegraphics[width=8.3cm]{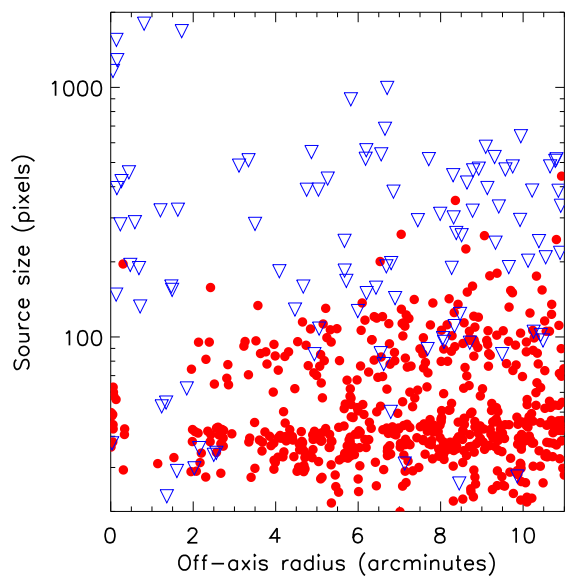}
\caption{The number of pixels enclosed by the XCS source ellipses
versus off-axis distance. The solid circles show point-like sources
(score $\le1$), while the open triangles are extended sources (score
$\ge3$). We note that the large on-axis sources are known, targeted,
nearby clusters of galaxies (see Section~\ref{known})\label{off}.}
\end{center}
\end{figure}

\begin{figure*} 
\begin{center}
\includegraphics[angle=0, width=0.8\textwidth]{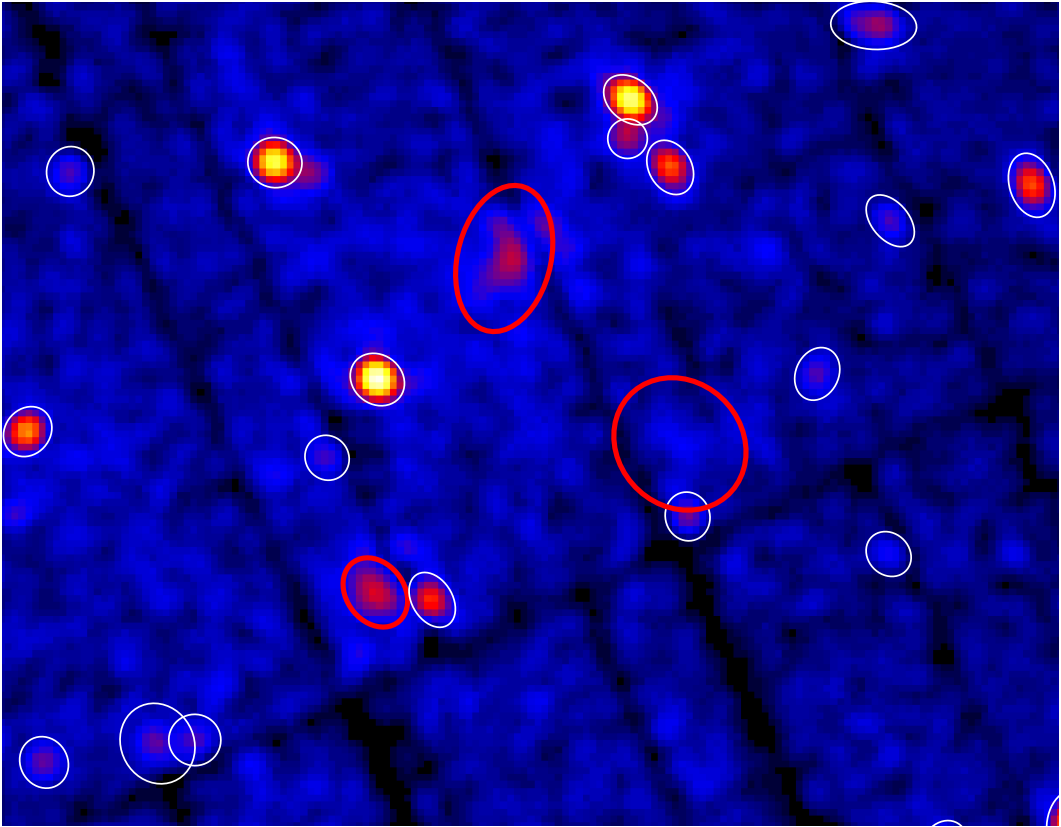}
\caption{Part of a (smoothed) XMM image showing three high-redshift
clusters redetected by XCS (thick rings); RDCS J0849+4452 ($z=1.265$; lower 
left),
RX J0848.7+4457 ($z=0.574$, SHARC cluster, top) and RDCS J0848+4453
($z=1.273$, lower right) (Hashimoto et al.~2002). The size of the ellipses 
reflects the
detected size of each object, showing that these high-redshift
clusters are resolved in this XMM data. The clusters are classified as
extended using our wavelet-ratio test, while the remaining objects in
the image are all classified as point-like. For
illustrative purposes we have shown the EPN image data, but the source
detection/parametrization was carried out in EMOS1, 
consistent with the other work presented herein. \label{RDCS}}
\end{center}
\end{figure*}

For each of the XMM/SDSS fields we used a combination of NED associations
and eyeballing to classify the sources found as either point-like, extended,
unknown or artifacts. When the various combinations of ratios were plotted
against each other it was found that there was differentiation between
these classifications, such that empirical cuts could be made to separate
extended and point-like sources. 
We use these
ratios as a crude extent classification, and extended sources can be
efficiently separated from point-like sources using the following
selection criteria:\\
\begin{center}
$R_{14} < 12.5\,\times R_{12}$ \quad or \quad $R_{14} < 4\,\times R_{13}$, \\ 
$R_{23}< 2$, \\ 
$R_{24} < 11$, \\ 
$R_{34} < 5$.\\
\end{center}
For each of these four criteria, we award each source a score of one,
so sources can have scores ranging from zero to four on this scale. 
To confirm that these cuts were
reasonable we calculated the histogram of extent scores found using them.
The distribution for sources classified as point-like showed that 90\% had
extent scores of 0 or 1, and similarly 90\% of the extended sources had
scores of 3 or 4.
For the remainder
of this paper, we use these scores as our classification of whether a
source is extended or not. In Figure~\ref{off} we present the size
(as the number of pixels enclosed by the source ellipse) of all
detected sources as a function of off-axis distance. This demonstrates
that there are two distinct populations of sources, regardless of
off-axis distance, and that our extent classification can successfully
separate them. 

To demonstrate the effectiveness of the method,
we show in Figure~\ref{RDCS} a 39 ksec XMM pointing (not in the SDSS
DR1 area) toward three previously-known high-redshift clusters,  all three of
which were classified as extended by our algorithm.  Of the
740 ($\theta<11$ arcminute) sources mentioned in Section~\ref{sode},
112 ($\simeq 15\%$) have extent scores $\ge3$. These 112 include
duplicate detections and sources falling in regions of the XMM field-of-view not
covered by SDSS DR1. We discuss the 99 unique extended sources with
SDSS DR1 coverage in Section~\ref{comb}.

We have tested the robustness of our extent algorithm using 59
duplicate detections. These sources lie in regions that overlap
between different XMM pointings.  We found that the source
classifications agreed in 53 cases (90\%). For the 6 sources that
disagreed, 5 of them are extremely bright point sources, with count
rates exceeding $1$ count per second. In these cases, the detailed structure
of the PSF is detected and the source appears extended rather than a
point source. If we exclude very bright sources, the
classifications agreed $98\%$ of the time.

We note that, subsequent to the work presented herein, we have
developed more sophisticated detection and classification
algorithms. We now use a modified version of WAVDETECT (Freeman et
al.~2002), rather than MR/1, to perform the wavelet
transforms. WAVDETECT has the advantage that we can use exposure map
information, something that is vital for the analysis of merged
(EMOS1+EMOS2+EPN) images. Furthermore, the extent classification now
uses models of the PSF provided in the CCFs [see Davidson et
al.~(in preparation) for more details].

\section{Optical Identifications using SDSS}

\label{s:sdss}

We have extracted from the SDSS DR1 website\footnote{See 
http://www.sdss.org/dr1/} all data available within a radius of 25
arcminutes around each of the 42 XMM pointing centers listed in
Table~\ref{pointings}. Eleven of the 42 have only partial SDSS
coverage, as described in the table comments. As X-ray
point sources (e.g.~stars and quasars) tend also to be optical point
sources, we have made use of the SDSS PHOTO flag {\em objtype} for
this study. This flag classifies optical sources as either point-like
({\em objtype}=6, e.g.~star, quasar), or extended ({\em
objtype}=3, e.g.~galaxy) and works well to ${\rm r}=20.5$. At
fainter magnitudes we have assumed everything is a galaxy (since
galaxies dominate the faint-end number counts). When SDSS spectra are
available, we can extract object redshifts and/or spectroscopic
classifications ({\em spec\_class}). Even when spectra are not
available, we can still make use of the results from the SDSS
spectroscopic target algorithms. These algorithms were designed to
select particular types of sources from the SDSS multi-color
photometry for spectroscopic follow-up. Many more targets are selected
than can be allocated optical fibres, e.g.~there are 16998 official
SDSS quasar targets in our study region alone.

In addition to the official SDSS target information, we use an extra
quasar classification flag ({\em QSOcol}) based on the colour--colour
diagrams of Richards et al.~(2001; see Figure~4 in their paper). From
these diagrams, it is clear that quasars predominately inhabit a
volume of colour--colour space as defined by: $-0.2< {\rm u}-{\rm g} <
0.7$, $-0.2< {\rm g}-{\rm r} < 0.6$, $-0.2< {\rm r}-{\rm i} <0.6$, and
$-0.2< {\rm i}- {\rm z}<0.5$.  Therefore, if an SDSS object satisfies
these colour cuts, we flag it as {\em QSOcol}$=1$.  Overall, 7443
objects were flagged in this way, of which only 1550 were already amongst the 
16998 official SDSS quasar targets. We use these {\em QSOcol}$=$1
objects in Sections~\ref{error}~\&~\ref{class}.

\begin{figure}
\includegraphics[width=8.3cm]{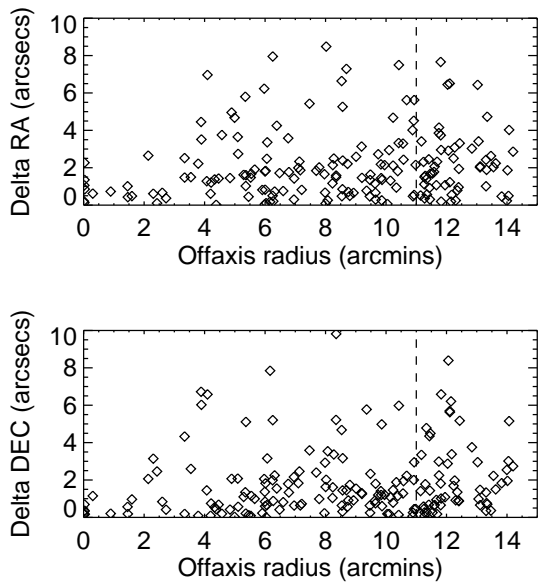}\\
\caption{The absolute scatter seen in the RA and DEC, as
a function of off--axis distance, for the 182 matches
between SDSS quasar candidates and XMM sources.}
\label{splitlevel}
\end{figure}

\subsection{Positional accuracy}
\label{error}

We have used the SDSS to derive the matching radius appropriate for
the optical follow-up of XCS sources. It is important that this radius
be as small as possible to cut down on erroneous matches. To do this,
we have made use of the fact that optical quasars should have a
point-like X-ray counterpart. We have compared XCS point sources
(i.e.~those with an extent score of $\le1$) that match, within an
initial search radius of 10 arcseconds, to a single SDSS point-like
object (i.e.~{\em objtype}$=$6).  Further, this single SDSS object
must have been spectroscopically classified by the SDSS as a quasar,
or be one of the 16998 official quasar candidate targets.

We find a total of 182 matches (no off-axis distance limit).  These
matches give an RMS of $3.8$ arcseconds for the total separation between
the X-ray and optical centroid. We also find an RMS of $\delta_{{\rm
RA}}=2.8$ arcseconds for the difference in Right Ascension and an RMS
of $\delta_{{\rm DEC}}=2.6$ arcseconds for the difference in
Declination for these matches. We show the scatter of $\delta_{{\rm
RA}}$ and $\delta_{{\rm DEC}}$ with off-axis distance for these 182
matches in Figure~\ref{splitlevel}, and the rms values in
Table~\ref{offaxis}.  We see an increase in the scatter
toward higher off-axis distances, which is to be expected due to the
increase in size of the XMM point-spread function with off-axis
distance. In the following, we therefore limit ourselves to the 103
matches at off-axis distances less than $\theta=11$ arcminutes.

\begin{table}
\caption{The error in arcseconds in matching sources as a function of off-axis 
distance.
\label{offaxis}}
\begin{tabular}{ccccc}\hline
Off-axis & Matches & rms $R$ & rms $\delta_{{\rm RA}}$ & rms $\delta_{{\rm 
DEC}}$\\ \hline
All & 182 & 3.8 & 2.8 & 2.6\\ 
$\theta<$8 arcmin & 73 & 3.4 & 2.6 & 2.3\\ 
$\theta<$11 arcmin & 125 & 3.7 & 2.9 & 2.4\\
$\theta>$11 arcmin & 57 & 4.0  & 2.8  & 2.9\\ \hline
\end{tabular}
\end{table}

\begin{figure}
\includegraphics[width=8.3cm]{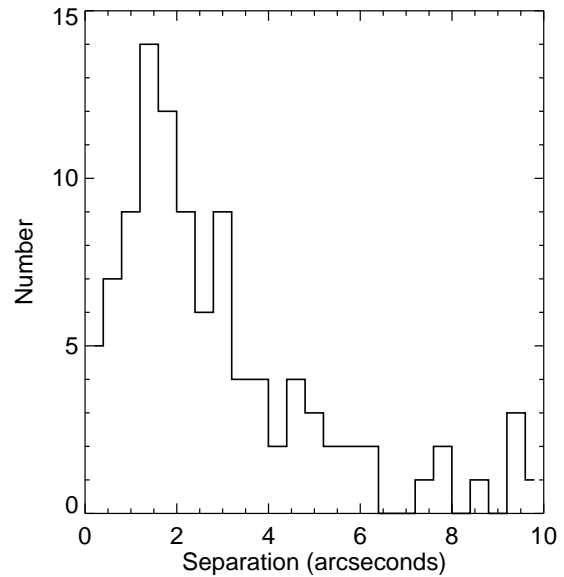}
\caption{The histogram of separations for the 103 SDSS quasars with XCS 
counterparts.}
\label{f:sepfit}
\end{figure}

\begin{figure}
\includegraphics[width=8.3cm]{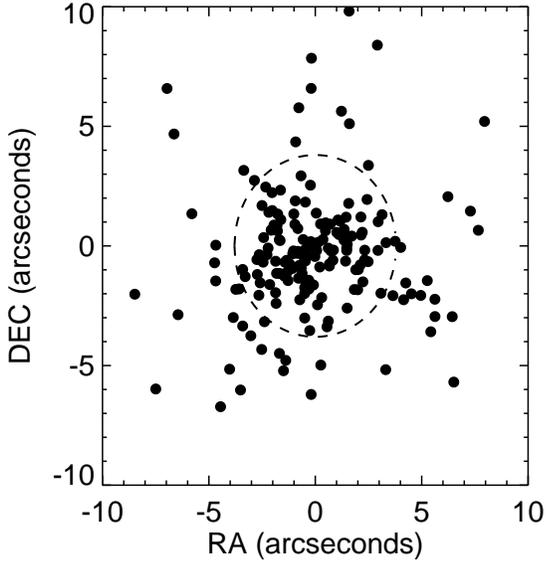}\\
\caption{The scatter in $\delta_{{\rm RA}}$ and $\delta_{{\rm DEC}}$ for 
103 SDSS quasars with XCS counterparts, with the dashed ring indicating the rms 
offset.}
\label{f:delta}
\end{figure}

In Figure~\ref{f:sepfit}, we present the distribution of absolute
separations between the optical and X-ray coordinates for 103 SDSS
quasars. If the errors in the RA and DEC are Gaussian distributions of
equal variance about zero, and the background sources negligible, the
separations should follow a $\chi^2_2$ distribution, whose variance
equals its mean.  However, those assumptions do not appear to be
valid, so we instead fit individual Gaussians to the observed
$\delta_{{\rm RA}}$ and $\delta_{{\rm DEC}}$ distributions for our 103
quasar candidates.  We find means of -0.4 and -0.6 arcseconds
respectively, with dispersions of 1.9 and 1.4 arcseconds.  The
distribution of separations is slightly elongated in the RA direction,
as can be seen in Figure~\ref{f:delta}.  The non-zero mean values
indicate that there is a systematic pointing offset, i.e.~errors in the 
boresight of each pointing, thus introducing a
systematic error in the coordinates of all sources in that pointing.

Furthermore, we note that the background sources are beginning to
dominate the source counts as we approach 10 arcsecond
separation. Therefore, we refined our analysis by including a linear
function in our fit to the distribution of separations to account for
the background contamination, i.e.~we fit a Gaussian plus a
linear function. The best-fit for the Gaussian part now has a mean of
1.8 arcseconds and dispersion of 1.2 arcseconds. Note that it is
the mean which estimates the typical positional error, while the
dispersion carries information about the width of the distribution.
We assume that matches within the 90\% confidence range of the
Gaussian part are likely to be true matches, obtaining a matching
radius of $1.8+1.65\times1.2 = 3.8$ arcseconds.

As a check of our determination of the matching radius, we split the
sample of 103 quasar candidates into bright and faint subsamples,
with a dividing flux of $1.2\times 10^{14} \, {\rm erg} \, {\rm
s}^{-1} \, {\rm cm}^{-1}$. We found that each subsample gave the same
90\% matching radius of 3.8 arcseconds using the prescription
above. We also checked duplicate (Section~\ref{extent}) X-ray sources
and found the mean separation between these sources to be consistent
with this matching radius. Furthermore, if we include a further 123
{\it QSOcol} sources (see above for definition) in our analysis, our
matching radius does not change. The size of the XCS matching radius
will be dominated by the positional errors in the X-ray data, as the
absolute astrometric calibration of the SDSS is accurate to less than
0.1 arcseconds (Pier et al.~2003) for point sources. Our matching
radius of $3.8$ arcseconds is similar to the 5
arcseconds suggested by Watson et al.~(2002), but is somewhat larger
than that found by Altieri (2002).  Altieri used matches with the USNO
catalogue to estimate individual XMM pointing offsets. He found a
systematic pointing offset of $\simeq0.3$ arcseconds, with a mean
offset of one arcsecond. Therefore, our $3.8$ arcseconds matching radius
likely includes two contributions, an error from the centroiding of
the sources and a systematic pointing offset.

\subsection{Optical overdensity}
\label{overden}

While the XCS is a serendipitous X-ray survey, with clusters to be discovered 
initially in XMM data, many such clusters will also be detectable in the optical 
as an overdensity in the projected galaxy distribution.
We have therefore calculated the
overdensity of SDSS galaxies around each of the X-ray sources in our
study region, by counting the number of galaxies
within an aperture of radius 40 arcseconds centered on the centroid of
the X-ray emission. This angular aperture was chosen to approximately
equal the angular size of distant clusters that could be seen in the
SDSS data; the typical core radius of a $z=0.35$ cluster (200kpc)
subtends an angle of 41 arcseconds on the sky. The area enclosed by
this radius ($\simeq 200$ pixels) is also in good agreement with the
size of extended XCS sources (Figure~\ref{off}).

We make a local background correction using the observed galaxy count
within an annulus centered on the X-ray source with an inner radius of
80 arcseconds and the outer radius equal to 100 arcseconds (scaled by
the area). This annulus corresponds to a radius of approximately half
the virial radius for a $z=0.35$ cluster, and thus provides a local
estimate of the projected galaxy density around each source. This
annulus will be contaminated by cluster members, but this effect will
be small. 

For each source, we calculate two galaxy overdensities, once imposing a
magnitude limit of ${\rm r}=22.2$ and once without a magnitude
limit. The former is the completeness limit of the SDSS photometric
survey (see A03), while the latter allows us to use the full dynamic
range of the SDSS photometry. For example, by not imposing a magnitude
limit we are effectively counting the number of galaxies detected in
any of the 5 SDSS passbands, as PHOTO (the SDSS photometric analysis
pipeline; Lupton et al.~2001) detects sources in all 5 passbands
separately and reports upper limits, if necessary, in the other
passbands if they are not detected. We stress again that for sources
brighter than ${\rm r}=20.5$ we use the SDSS PHOTO {\em objtype} flag
to identify galaxies, while below this limit, we simply call all
objects galaxies. We also use any object spectroscopically confirmed
to be a galaxy or listed as an official galaxy target. By using a
local background correction, we guard against fluctuations both in the
completeness of the SDSS photometric survey and in the star--galaxy
separation at faint magnitudes.

\begin{figure}
\centering
\includegraphics[width=8.3cm]{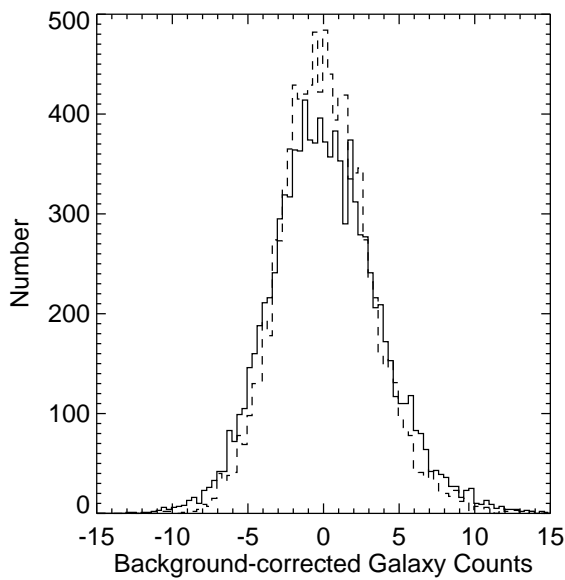}\\
\caption{The distribution of background corrected galaxy counts from both the
no magnitude limit case (solid line) and the $r=22.2$ limited case
(dashed line).}
\label{prob}
\end{figure}

To determine the statistical significance of any observed overdensity
of galaxies around these X-ray sources, we construct the distribution
of counts one would expect from random. This was achieved by placing
the 40 arcsecond radius aperture on 10,000 randomly selected SDSS
sources and making the same local background correction. We only
consider SDSS sources from the XMM regions that have full SDSS
coverage (see Table~\ref{pointings}) and restrict our analysis to SDSS
sources between 3 and 11 arcminutes from the centre of those
pointings.  We note that there are more than 10,000 SDSS sources that
satisfy this criteria (mostly distant galaxies and faint stars) and
their angular clustering is weak, less than 1\% angular fluctuations
within a 40 arcsecond aperture (Connolly et al.~2002), especially for
the no-magnitude count distribution. We show the two count
distributions in Figure~\ref{prob}. As expected the distribution for
the no magnitude limit case distribution is broader.  The
distributions are well fit by a Gaussian with a mean and dispersion of
$-0.11$ and 3.39 respectively for the no magnitude limit case and
$-0.074$ and 2.86 for the $r=22.2$ limited case. We re-fit the
Gaussian using only the count data between -5 and 5, but found
consistent means and sigmas. We tried other methods for constructing
the count distributions, e.g.~randomly placing the apertures
within the SDSS area, and found the results to be consistent.

We have used the distributions presented in Figure~\ref{prob} to
define a threshold above which an observed SDSS overdensity is
considered to be statistically significant. By design, such
overdensities are ideal candidates for clusters of
galaxies. To define this threshold, we have used the False Discovery
Rate (FDR) statistical methodology as outlined in detail by Miller et
al.~(2001) and Hopkins et al.~(2003). It has been applied to SDSS
cluster selection in Miller et al.~(2004). Briefly, FDR controls the
ratio of the number of false detections over the total number of
detections, and therefore is a more scientifically-meaningful quantity
than selecting a fixed threshold based on a multiple of sigma, 
e.g.~using a 3-sigma threshold. FDR is adaptive in that it changes
based on the number of tests performed, while a standard sigma-based
threshold is not adaptive and thus leads to many more false detections
as the number of tests grows large.  In summary, FDR is ideal for our
problem, especially as it naturally accommodates correlated data which
is the case here (see Miller et al.~2001).

The key parameter of FDR is $\alpha$, the ratio of false to true
detections one is willing to tolerate. Here, we have selected $\alpha$
to be 0.25, i.e.~we will allow up to 25\% of our selected
overdensities to be potentially false.  Computationally, FDR is simple
to implement and involves determining the probability that the
observed SDSS count around each of our X-ray sources is drawn from the
count distributions in Figure~\ref{prob}. These probabilities are then
ranked and a threshold calculated using the methodology given
in Miller et al.~(2001). This was performed separately for both
distributions featured in Figure~\ref{prob}. Using $\alpha=0.25$ we
selected 39 candidate overdensities from each distributions, i.e.~in both cases 
39 of the 690 X-ray sources were assessed to lie
in statistically-significant galaxy overdensities. There is
significant overlap between the two lists, with 30 candidates in
common giving 48 candidates in total. We discuss these 48 galaxy
overdensities further in Section~\ref{comb}. For reference, the
calculated FDR thresholds corresponds to a threshold of 2.20 sigma (an
overdensity above 7.3) for the no magnitude limit distribution and
2.24 sigma (above 6.3) for the ${\rm r}<22.2$ count distribution.

\subsection{Photometric redshifts}
\label{photoz}

\begin{figure} 
\begin{center}
\includegraphics[angle=0, width=3.5in]{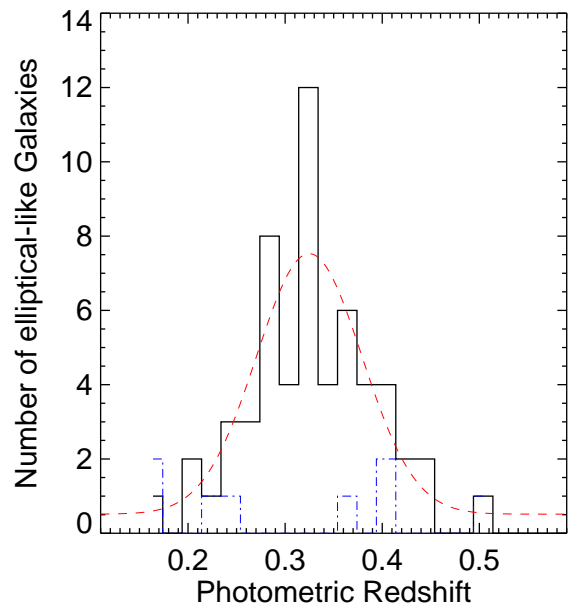}
\caption{The photometric redshift histogram of cluster RXJ0256 
(black line).  The blue/dashed histogram represents the photometric
redshift histogram for a randomly chosen location 1.3 degrees west of
the cluster centroid.  This demonstrates the significance of the peak
seen in the black cluster histogram, which contains 60 sources
compared to only 8 sources in the blue/dashed
histogram.\label{f:photoz}}
\end{center}
\end{figure}

\begin{table*}
\caption{The classification of our 637 XMM sources.}
\label{digits}
\begin{tabular}{lrl} \hline
Class & Number & Comments \\ \hline

0100&   203  &   Not extended, one point-like SDSS match\\
0101&    11  &    Not extended, two SDSS matches (at least one 
is point-like)\\
0000&   296  &   Not extended, no point-like SDSS match (could have 
an SDSS galaxy match within $3.8''$)\\
0001&    3   &    Not extended, two SDSS matches (neither are 
point-like)\\
1000&   70   &   Extended, no point-like SDSS match (could have an 
SDSS galaxy match within $3.8''$ )\\
1101&    2   &    Extended, two SDSS matches (at least one is 
point-like)\\
1100&    11  &    Extended, one point-like SDSS match\\
1010&   16   &    Extended, no point-like SDSS match, galaxy 
overdensity\\
0010&   15   &   Not extended, no point-like SDSS match, galaxy 
overdensity (could have 
an SDSS galaxy match within $3.8''$)\\
0110&   8    &   Not extended, one point-like SDSS match, and galaxy 
overdensity\\
0011&    2   &    Not extended, two SDSS matches (neither are 
point-like), galaxy overdensity\\
0111&    0   &    Not extended, two SDSS matches (at least one is 
point-like), galaxy overdensity\\
1111&    0   &    Extended, two SDSS matches (at least one is 
point-like), galaxy overdensity \\
1110&   0   &    Extended, one SDSS point-like match, galaxy 
overdensity\\
1011&   0    &   Extended, two SDSS matches (neither are point-like), 
galaxy overdensity\\
1001&   0   &    Extended, two SDSS  matches (neither are point-like)\\ 
\hline
\end{tabular}
\end{table*}

When SDSS data are available for a particular XCS cluster candidate, it
may be possible to determine the cluster redshift without the need for
additional dedicated spectroscopic follow-up. We have investigated how
well we can recover the spectroscopic redshift for the $z>0.1$
clusters using photometric
redshifts. Photometric redshift estimates are available for all
galaxies in the SDSS DR1 (see Csabai et al.~2003; Budav\'{a}ri et al.~2003).  
Briefly, the SDSS DR1 database contains an
estimate of both the redshift and spectral type ($t$), with errors,
for each galaxy. For each $z_{{\rm spec}}>0.1$ cluster identified, we extracted 
these data from the DR1 in an
aperture of radius 40 arcseconds (i.e.~the same as was used for the
optical overdensity measurements) centered on the X-ray centroid. We
then excluded any galaxies with $z_{\rm photo}<0.05$, with a
photometric redshift error of greater than the measured photometric
redshift, or with a photometric spectral classification of $t\ge 0$.
This last criterion ensures we are only considering
``elliptical-like'' galaxies which are likely to be cluster members
(see Budav\'{a}ri et al.~2003 for a discussion of the $t$ parameter). We
then fit the distribution of these remaining photometric redshifts
with a Gaussian plus a constant (e.g.~Figure~\ref{f:photoz}) to
accommodate outliers in redshift (mostly at redshifts greater than the
cluster). 

Alternative approaches would be
to use the E/S0 ridgeline (e.g.~Gal et al.~2000; Gladders \&
Yee 2000) or a matched filter technique. The latter technique has
recently been applied to SDSS data by Schuecker et al.~(2004). Based on 60 
clusters detected using a joint RASS/SDSS search,
they found an average redshift offset of $\sigma_z=0.03$, or 20\% at
their typical survey redshift ($z\simeq0.15$).

\section{Matched sources and cluster candidates}
\label{comb}

We now return to X-ray source identifications using SDSS data.  We
concentrate on XCS sources detected in one or more of the 42 pointings
listed in Table~\ref{pointings} at off-axis distances of $\theta<11$
arcminutes.  We detected 740 such sources, 690 of which also overlap
with the SDSS DR1.  Of these, 53 are duplicate detections, leaving 637
unique sources. Amongst these 637, 520 are point-like (extent
score $\leq 1$) and 99 are extended (extent score $\geq 3$). The
remaining 18 have an ambiguous classification (extent score $=2$); henceforth we 
will include them under the description `point-like'.

\subsection{Source classification}
\label{class}

In Table~\ref{digits} we present the results of matching the 637 XMM
sources with the SDSS DR1 data, using a matching radius of 3.8
arcseconds (Section~\ref{error}) centered on the X-ray centroid. We
also present the results of our search for galaxy densities
(Section~\ref{overden}). To understand the SDSS--XMM match-ups, we use
the following four binary criteria (0$=$no, 1$=$yes) when classifying
all the XMM sources:
\begin{itemize}
\item Is the X-ray source extended (extent $\ge 3$)?
\item Is there at least one SDSS point source ({\em objtype}$=6 $)
present within the 3.8 matching radius? 
\item Is there a significant overdensity of SDSS galaxies? 
\item Are there two SDSS sources within the 3.8 arcsecond matching
radius? For this test, we include all SDSS objects i.e.~also
galaxies ({\em objtype}$=3 $).
\end{itemize}
Each source has a 4-bit identification, e.g.~0011, 1010 etc., providing sixteen 
possible combinations.
The ordering of the bits is as given in the list above.
The breakdown of classifications for all 637 unique XMM sources in given in
Table~\ref{digits}.

We find 214 point-like X-ray sources that match at least one
point-like SDSS source ($0100+0101$): 55 of these SDSS point-like
matches are stars, 159 are quasars. Therefore, we have been able to
provide immediate identifications for 34\% of all sources and can
eliminate them from a cluster candidate list without any need for
dedicated optical follow-up.  We provide no further discussion herein
of these 214 sources, beyond noting that such merged XMM/SDSS point-source 
catalogues will be a powerful resource for AGN/quasar science.

We find 299 (47\% of the total) point-like X-ray sources that have no
point-like SDSS counterpart ($0000+0001$). For these, we cannot make
an immediate classification based on SDSS data alone. Most of them
will be optically-faint stars or active galaxies/quasars and are not
of an immediate relevance to the XCS science goals.  However, we note
the three $0001$ cases, where an X-ray point source lies within
$3.8$ arcseconds of two SDSS galaxies. For these, we may be seeing a few
brightest cluster galaxies and further optical follow-up is
merited. By contrast, we do not plan to pursue dedicated optical
follow-up of the 296 objects with the $0000$ flag.  A small fraction
of them may be clusters or groups. However, they have neither been
flagged as galaxy overdensities in the SDSS or as extended by our
X-ray pipeline. The XCS extent algorithm works even at high redshifts
(see Figure~\ref{RDCS}), but only given a high enough signal-to-noise.
To simplify the XCS selection function, we are adopting a conservative
signal-to-noise threshold (see Romer et al.~2001). Any
clusters in the $0000$ category must be low signal-to-noise detections
and would not be suitable for inclusion in the XCS cluster catalogue.

\begin{table*}
\caption{Known clusters in our XMM pointings. Those above the line
were target clusters, listed in Table~\ref{pointings}. The fluxes are
in units of $10^{-12}$ erg s$^{-1}$ cm$^{-2}$.  We estimated
photometric redshifts only for clusters with $z_{{\rm spec}} > 0.1$.}
\label{clusters}
\begin{tabular}{lccccccccc} \hline
Cluster &  Extent &    Major axis &  Minor axis &  Off-axis angle & Counts &      
    Flux & SDSS & $z_{{\rm photo}}$ & $z_{{\rm spec}}$\\
       &  Score  &    (arcmin)   &  (arcmin)   &  (arcmin) & (s$^{-1}$) &  & 
overdensity 
& \\ \hline

RXCJ2337.6+0016/Abell 2631   &  4  &    2.42  &    1.71  &  0.81   &  0.212  &   
1.094 & yes & $0.26\pm0.02$ & 0.273 \\
RX J0256.5+0006               &  4  &    1.13  &    0.85  &  0.26   &  0.035 &   
0.193 & yes & $0.32\pm0.05$ & 0.36\\
RX J0256.5+0006  (subclump)                     &  4  &    0.75  &    0.58  &  
0.68   &  
0.008  &   0.043 & yes & $0.32\pm0.05$ & 0.36\\
Abell 85                              &  4      &    1.15  &    1.12  &  6.19   
&  0.022  &   0.111 & yes &                 & 0.044\\
Abell 267                     &  4  &    1.95  &    1.51  &  0.16   &  0.242  &   
1.243 &     & $0.23\pm0.03$ & 0.23\\
Abell 773                           &  4  &    2.06  &    1.73  &  0.14   &  
0.298  &   1.457 & yes & $0.21\pm0.02$ & 0.217\\
Abell 1689                    &  4  &    1.69  &    1.58  &  0.05   &  1.00  &   
4.909 &     & $0.19\pm0.04$ & 0.1832\\
MKW4                          &  4  &    0.85  &    0.78  &  0.58   &  0.251  &   
1.257 & yes &                 & 0.02\\
Abell 1650                    &  4  &    2.31  &    1.67  &  1.72   &  0.853  &   
4.281 &     &                 & 0.0845\\
MS1137.5+6625                 &  4  &    1.06  &    0.99  &  0.44   &  0.022  &   
0.107 & yes &                 & 0.7820\\
Abell 2670                    &  4  &    0.91  &    0.81  &  1.62   &  0.038  &   
0.196 &     &                 & 0.0762\\
Abell 2670 (subclump)                    &  4  &    0.38  &    0.33  &  1.34   & 
 0.013  &   
0.065 &     &                 & 0.0762\\
ZwCl 1021.0+0426              &  4  &    0.88  &    0.74  &  0.23   &  0.489  &   
2.461 & yes & $0.27\pm0.03$ & 0.291 \\ 
\hline
VMF98-116                     &  4  &    1.75  &    1.18  &  5.82   &  0.017  &   
0.083 & yes & $0.36\pm0.02$ & 0.409 \\
VMF98-021                       &  4  &    0.79  &    0.60  &  10.55  &  0.034  
& 
  0.176 & yes & $0.35\pm0.03$ & 0.386 \\
SDSS CE J010.717058             &  4  &    1.15  &    0.97  &  10.65  & 0.004  & 
  0.018 & yes &                 &  \\
\hline\end{tabular}
\end{table*}

More interesting to the XCS are the 70 extended X-ray sources (category 1000) 
that have no SDSS counterpart (point-like or
galaxy). Of these 70, 44 have the highest possible extent score
(4). These are excellent distant cluster candidates, with galaxy
populations too faint to be detected in the SDSS. These are our
top-priority targets for dedicated optical follow-up. An additional 13
extended sources are not associated with a galaxy overdensity, but
match to at least one SDSS point source ($1101+1100$). These require
optical follow-up; optical spectroscopy of the SDSS point sources
would demonstrate if they are likely X-ray emitters (e.g.~emission
line galaxies). The $1101+1100$ objects might also merit X-ray
follow-up using Chandra.  Chandra has better spatial resolution than
XMM and would elucidate any cases where multiple point sources have
been blended together by XMM to mimic an extended source, or where
genuine extended cluster emission is contaminated by point-source
emission. Therefore, even though the SDSS cannot provide a definitive
source identification in these cases, we can use it to highlight
objects that might be contaminating our cluster candidate list and
adjust our follow-up strategy accordingly.

The remaining classes of X-ray sources ($1010$, $0010$, $0110$,
$0011$) all have an associated SDSS galaxy overdensity, 41 in total.
Based on our FDR threshold selection, we expect 75\% of these to be
clusters, i.e.~true physical associations of galaxies. Such clusters
may or may not be detected as extended sources by XMM, depending on
their redshift and mass and on the sensitivity of the particular XMM
pointing they fall in. We discuss these 41 objects further in
Sections~\ref{known} and \ref{new}.
 
\subsection{Known clusters}
\label{known}

We have detected 14 previously-catalogued clusters as XCS
sources. Three are serendipitous rediscoveries, the rest were the
intended target of their respective XMM pointing. 
All clusters were detected as extended (extent score of 4), 10 were
also flagged as being associated with SDSS galaxy enhancements.  The
clusters are listed in Table~\ref{clusters}, together with information
about the extent score, source size, off-axis distance, count rate, flux,
galaxy overdensity and redshift. We note that RX J0256.5+0006 (Obs
ID 0056020301) and Abell 2670 (Obs ID 0108460301) are listed twice,
because in both cases our software has detected an extended
subcomponent to the main cluster. The RX J0256.5+0006 subclump has
been noted during an earlier analysis of the XMM image (Majerowicz et
al.~2004). We further note that Abell 85, despite being the pointing
target, was detected at a significant off-axis distance. Abell 85 is so
large, compared to the adopted wavelet kernels, that our source detection
software has only recovered part of the total flux from this system.

The XCS is a serendipitous survey. Therefore, unless known clusters
are genuine serendipitous redetections, they will be excluded from the
XCS cluster catalogue used for cosmological studies. However, our
examination of these 14 known clusters has demonstrated several
important features of the XCS software pipeline; it can detect
clusters out to $z=0.782$ (MS1137.5+6625) as extended objects, it can
cope with situations where there is cluster substructure (RX
J0256.5+0006 and Abell 2670) or where there are bright point sources
nearby (MS1137.5+6625), and it works even far off-axis (VMF98-021 and
SDSS CE J010.717058 were both detected at $\theta>10$ arcminutes). We
note that cluster VMF98-173 ($z=0.112$, Mullis et al.~2003) was also
detected, with an extent score of 4, at an off-axis distance
$\theta>11$ arcminutes in pointing 0060370101.  In summary, our XMM
source detection algorithm has recovered all previously-known distant
X-ray clusters within our surveyed area, and all with an extent score
of 4.

We have also been able to demonstrate the potential of the SDSS
archive to provide galaxy data for medium-redshift clusters. Ten of
the 14 clusters ($0.044<z<0.782$) were associated, by our FDR
methodology, with SDSS galaxy enhancements. The SDSS is not expected
to be of particular use for high-redshift cluster follow-up ($z>0.5$,
Schuecker et al.~2004), but occasionally we will pick
up rich systems as galaxy enhancements. For example, MS1137.5+6625, a
very bright EMSS cluster (Gioia et al.~1990) at $z=0.782$, was
detected as an overdensity of galaxies in the SDSS. In the no
magnitude limit search, 7.8 (background-corrected) galaxies were found
within the 40 arcsecond aperture (a 2.3 sigma excess). However, in
general we will not be able to rely on SDSS data as the primary
resource for cluster selection.  Rather we will use X-ray extent to
separate clusters from the general X-ray source population.

We have used the SDSS data to determine photometric redshifts for the clusters 
with $z>0.1$, with the mean and sigma of the fitted Gaussian presented in
Table~\ref{clusters}.
Acceptable fits were found for all except the highest-redshift cluster
(MS1137.5, $z_{{\rm spec}}=0.782$).
Overall, all our photometric redshifts are in good agreement with the
spectroscopic redshifts and, except in the case of VMF98-116, they are
within the quoted one-sigma errors on the photometric cluster
redshift.  For VMF98-116, the photometric redshift is within two-sigma
of the published spectroscopic redshift (Mullis et al.~2003).  We note
that one galaxy in the 40 arcsecond aperture around VMF98-116 has a
measured SDSS spectroscopic redshift of $z_{{\rm spec}}=0.412$,
i.e.~very close to the published cluster redshift (the SDSS
photometric redshift of this particular galaxy is $z_{{\rm
photo}}=0.43$). The photometric redshift histogram of VMF98-116 is
bimodal with a secondary peak at $z_{{\rm photo}} \simeq 0.42$,
suggesting this cluster is undergoing a merger.  We also highlight the
case of RXJ0256. This cluster is known to be undergoing a merging
event (Majerowicz et al.~2004). The photometric redshift histogram
(Figure~\ref{f:photoz}) is broad and has two secondary peaks.  The
average offset between the photometric and spectroscopic redshifts for
the eight clusters is 6.4\%. Excluding RXJ0256 and VMF98-116, with
offsets of 11\% and 12\% respectively, reduces the average offset to
4.7\%. We note that in all but one case, Abell 1689, the photometric
redshift was lower than the spectroscopic redshift, suggesting a
systematic bias in the technique.

\subsection{X-ray properties of SDSS galaxy overdensities}
\label{new}

Of the 41 X-ray sources statistically associated with SDSS galaxy
overdensities, 16 are also classified as extended (category 1010 in
Table~\ref{digits}).  These are very likely to be true X-ray clusters,
although we would not expect them to be at particularly high redshifts
given the magnitude limit of the SDSS.  The remaining 25 overdensities
are associated with X-ray point sources; 15 in category 0010, 8 in
category 0110 and 2 in category 0011.  We re-emphasize that we expect
25\% (roughly 10) of the 41 FDR flagged overdensities to be false
discoveries.  Without further follow-up we cannot determine which
overdensities are false; however it is fair to speculate that those in
category 0110 (X-ray point source, SDSS point source match) are more
likely to be false than those in category 1010 (extended emission, no
SDSS point source match).  Of the 16 category 1010 sources, 11 are
redetections of 10 known clusters; 7 target clusters (one resolved
into two known components) and 3 serendipitous detections (see
Section~\ref{known} and Table~\ref{clusters}). These were all detected
with an extent score of 4.  The remaining 5 sources in the 1010
category are prime cluster candidates, two with an extent score of 4 and
three with an extent score of 3.  Using the methodology outlined in
Section~\ref{photoz}, we have determined $z_{\rm photo}$ estimates for
these candidates to be $0.32 \pm 0.01$, $0.46 \pm 0.03$, $0.35 \pm
0.04$, $0.27 \pm 0.03$ and $0.45 \pm 0.03$ respectively.  For one of
them (at 11h 50m 11.40s 01d 38m 45.2s J2000), there is also a single
SDSS galaxy spectroscopic redshift available, $z_{{\rm
spec}}=0.451$. We plan to make follow-up observations of these five
cluster candidates to secure identifications and measure spectroscopic
redshifts.

We now turn to the 25 overdensities associated with X-ray point
sources. Clusters of galaxies should be detected as extended sources
by XMM, so the nature of these 25 sources is intriguing. Some or all
of these sources could be low signal-to-noise detections of clusters,
for which the extent classification is unreliable. We will
re-examine these objects in the merged (EMOS1+EMOS2+EPN) images
(Section~\ref{xmm}) and using the new extent algorithm
(Section~\ref{extent}).  However, as mentioned above, approximately 10 of the
overdensities will be false discoveries. In fact, there are 8 sources
(category 0110) that have been matched to an SDSS point source. In
these cases the SDSS object, rather than a cluster, is likely to be
the source of the X-ray emission (see above for the discussion of the
214 category $0100+0101$ sources). In another 9 cases, an SDSS galaxy
was found within the $3.8$ arcsecond matching radius (7 sources in the 0010
category and both in the 0011 category). These 9 could be cases where
there is a genuine physical association of galaxies, but where the
X-ray emission is coming predominantly from an active galaxy (rather
than from an extended intracluster medium).

\section{Discussion}

In this paper we have presented preliminary results from the XMM
Cluster Survey [XCS, Romer et al.~(2001)], restricting our analysis
to those areas covered by the first data release of the SDSS. Our
principal goal has been to investigate the extent to which SDSS data
can assist with the massive follow-up program required to compile a
galaxy cluster catalogue. In this pilot study we have explored 1121
X-ray sources identified in 42 pointings from the XMM data
archive. The $\log N$--$\log S$ relation indicates a survey
completeness down to a flux limit around $1\times10^{-14} \, {\rm erg}
\, {\rm s}^{-1} \, {\rm cm}^{-2}$.  By using only the central 11
arcminutes of the XMM frames, and removing duplicate sources from
overlapping pointings, we have reduced this to a set of 637 unique sources,
of which 99 are identified as extended (extent score $\ge 3$) by our
wavelet algorithm. We expect that all clusters detected at sufficient
signal-to-noise, including those at high redshift, will be resolved by
the XMM instrument. We have demonstrated this by redetecting as
extended two previously-catalogued clusters at $z=1.27$. 

We have used SDSS quasars to determine a 3.8 arcsecond matching radius
that is appropriate for XMM follow-up.  We have also used the False
Discovery Rate (FDR) methodology to find SDSS galaxy overdensities
within a radius of 40 arcseconds of the X-ray sources.

Sixteen sources are identified both as extended in the X-rays and
coincident with an SDSS galaxy overdensity, and are therefore very
strong nearby cluster candidates. In fact 11 of them correspond to known
clusters (mainly the targets of the XMM pointings), including a
distant cluster at $z= 0.78$.  The remaining 5 are new cluster
candidates with photometric redshifts in the range
of $0.25<z_{\rm photo}<0.5$.  We have determined photometric redshifts for
these 16 sources using the galaxy photometric redshift information in the
SDSS archive. The offset between the measured and estimated redshifts
is $\simeq 6\%$ overall, but is significantly larger ($\simeq 11\%$)
for merging systems. This demonstrates that additional (to SDSS)
spectroscopic follow-up will be necessary before we can accurately
model the XCS survey selection function. This requirement applies also
to other cluster surveys (X-ray and Sunyaev--Zel'dovich) that require
redshift data before cosmological parameters can be estimated.

We find that only relatively nearby clusters, most of which are known
already, are detected by both indicators. The best candidates for
distant clusters are the 83 sources detected as extended X-ray sources
but with no corresponding SDSS overdensity, as the SDSS selection
function falls off rapidly beyond $z \sim 0.5$ (Schuecker et
al.~2004). In particular, 70 of these sources have no SDSS counterpart
at all; the majority (44) were assigned the highest possible
extent score. The remaining 13 are coincident with an SDSS point
source, and may be clusters but possibly with some contamination.

There are 25 sources which are point-like in the X-rays but which are
coincident with SDSS overdensities. Some will be accidental
superpositions, but the level of these is controlled by the False
Discovery Rate and the majority (75\%) will be real. The precise
nature of these sources is unclear; some could be clusters where an
embedded, or projected, point source is dominating the X-ray emission.

The remaining sources are point-like in the X-ray, and either have no
SDSS match at all or match an SDSS point source. These sources are
unlikely to be clusters and are low priority for follow-up as XCS
clusters. We note however that we have a sizeable sample of SDSS point
sources matched to X-ray point sources; most of these are likely to be
AGN/quasars and represent a powerful resource for studying that
population.

In summary, use of the SDSS does not replace the need for follow-up of
XCS cluster candidates, but can play a significant role in optimizing
the follow-up strategy. The simplest strategy would be to focus only
on extended X-ray sources, in which case the SDSS can split the sample
into nearby clusters detected as SDSS overdensities, distant cluster
candidates which may have point source contamination, and distant
cluster candidates with no associated point source. Additionally, the
SDSS can provide some further candidates (25 in our case) where
point-like X-ray sources are found to be coincident with galaxy
overdensities.

Finally, we note that the recent second data release from the SDSS
(Abazajian et al.~2004), plus the ongoing growth of the XMM data
archive, has led to a multiplication in overlap area by a factor of
nearly four as compared to the area studied in this paper. We will
analyze this increased overlap area in future work.

\section*{Acknowledgments}

We acknowledge financial support from the Royal Astronomical Society's
Hosie Bequest (KRL, MD), PPARC (KRL, ARL, STK, RGM), NASA LTSA award
NAG-11634 (AKR, RCN, KS, MD, PTPV), the NASA XMM program (KS), the
Institute for Astronomy, University of Edinburgh (MD), Carnegie Mellon
University (KS, AKR), the Leverhulme Trust (ARL), and NSF grant
AST-0205960 (MJW). We thank Peter Freeman for useful discussions
concerning source detection, Shane Zabel for advice regarding the XCS
data pipeline, and William Chase for coding and data archiving
assistance.  KRL thanks Joao Magueijo and Mike and Rosemary Land for
their support and encouragement.

\bsp

\end{document}